# Incorporation of New Information into an Approximate Hamiltonian

**C. P. Viazminsky**
Department of Physics
University of Aleppo, Syria

**Sohail Baza**
Department of Physics
University of Aleppo, Syria

**Abstract**
Additional information about the eigenvalues and eigenvectors of a physical system demands extension of the effective Hamiltonian in use. In this work we extend the effective Hamiltonian that describes partially a physical system so that the new Hamiltonian comprises, in addition to the information in the old Hamiltonian, new information, available by means of experiment or theory. A simple expression of the enlarged Hamiltonian, which does not involve matrix inversion, is obtained. It is also shown that the Lee-Suzuki transformation effectively put the initial Hamiltonian in a diagonal block form.

## 1. Introduction

It is well known that Schrodinger equation cannot be solved analytically except in limited simple cases. This calls upon approximation methods, of which perturbation method is most common [1,2], to find approximate solutions of this equation. The perturbation method hinges on ignoring a part of the Hamiltonian $H$, called the perturbed part, so that the resulting equation is solvable analytically. The full Hamiltonian $H$ is described as perturbed, whereas the simplified one is described as unperturbed. The unperturbed equation is solved analytically, and the corrections that take into account the ignored part are calculated. The set of eigenvectors $\{e_i\}_1^\infty$ of the unperturbed Hamiltonian is taken as a basis in the infinite dimensional Hilbert space of wave functions, denoted by $H_\infty$. Observables pertaining to the system are represented by Hermitian matrices in terms of this basis. Unless these matrices are given by recurrence formulae, we have to be contented with finite matrix approximation, which implies truncating the infinite basis at some sufficiently large term $N$ [12]. The space generated by the truncated basis $\{e_\mu\}_1^N$ -denoted by $H$ and called the truncated space- will hopefully contain good approximation of all states of interest to the problem we consider.

It must be noted that, whenever the eigenvalue problem is to be solved numerically, which is usually the case in physically interesting problems, truncation is inevitable task. We also note that truncating an infinite basis by a finite one with a sufficiently large number of basis elements is justified by the fact that the sequence $(e_N)$ tends weakly to zero as $N$ tends to infinity. This means that for every wave function $\psi \in H_\infty$ the sequence of numbers $(<e_N | \psi>)$ tends to zero as $N$ tends to infinity [3,4]. Alternatively, an upper cutoff, N, can be safely applied without seriously changing the low-lying properties [12].

Supposing that the truncated space H is good enough to replace the full Hilbert space, the full Hamiltonian can be represented by a $(N \times N)$ matrix that acts in the truncated space. It must be emphasized that truncation is dictated by the computation process and is independent of the approximation method employed. It is quite likely that the approximation method may yield good estimates $\{E_1,...,E_d\}$ of the first $d$ levels, and that the accuracy decreases as the number $n$ of the level increases, so that it becomes unacceptable for $n > d$ ($n$ is still much smaller than N). Assuming that some experimental facts or theoretical considerations assures that the numbers $\{E_{d+1},...,E_n\}$ are good estimates of the next ($n$-$d$) levels, then the question arises as how to construct a new approximate Hamiltonian whose spectrum is $\{E_1,...,E_n\}$.

A second approach in approximation, namely effective Hamiltonian methods [5-12], is common in nuclear and molecular physics. This approach is based on constructing a Hamiltonian $H_{11}$ in a $d$-dimensional subspace of the truncated space H that possesses $d$ eigenvalues of the full Hamiltonian. The question also arises how to incorporate in this Hamiltonian additional data based on experiment or theory. Our goal in this work is to study the problem associated with this question in an abstract sense, and give a rigorous answer of the question we just posed.

## 2. Algebraic Considerations

Let $\Pi$ be a d-dimensional Hilbert space. $\Pi$ is then isomorphic to $C^n$. The natural embedding

$$(x_1,...,x_d)^t \in \Pi \to \psi = (x_1,...,x_d,0,0,...,0)^t \in H_n$$

of the space $\Pi$ in the Hilbert space $H_n$ ($n > d$) is an isomorphism between $\Pi$ and a $d$-dimensional subspace of the Hilbert space $H_n$. Through this isomorphism we may identify every vector $\alpha_i \in \Pi$ with a vector $\psi_i \equiv \begin{pmatrix} \alpha_i \\ 0 \end{pmatrix} \in H_n$. Let $L_{11}$ be a linear operator in the Hilbert space $\Pi$. If $\{e_\mu\}_{\mu=1}^n$ is a basis in $H_n$, then the operator $L_{11}$, which is represented in the basis $\{e_i\}_1^d$ by a $(d \times d)$ matrix $L_{11}$, may also be viewed as an operator in $H_n$ with the corresponding matrix representation

$$l = \begin{bmatrix} L_{11} & 0 \\ 0 & 0_{n-d} \end{bmatrix}. \quad (1)$$

Thus we have

$$l \begin{bmatrix} \alpha_i \\ 0 \end{bmatrix} = \begin{bmatrix} L_{11}\alpha_i \\ 0 \end{bmatrix} \quad \forall \alpha_i \in \Pi \quad (2)$$

Assume now that $L_{11}$ possesses the discrete spectrum $sp(L_{11}) = \{E_1,...,E_d\}$ to which the linearly independent eigenvectors $\{\alpha_1,...,\alpha_d\}$ belong. And let $\{E_{d+1},...,E_n\}$ be a set of numbers $\{\psi_{d+1},...,\psi_n\}$, be a set of n-vectors such that the set of n-vectors

$$\Psi = \{\psi_1 = \begin{pmatrix} \alpha_1 \\ 0 \end{pmatrix}, \ldots, \psi_d = \begin{pmatrix} \alpha_d \\ 0 \end{pmatrix}, \psi_{d+1}, \ldots, \psi_n\}$$

is linearly independent. The problem which we consider is how to construct an operator

$$L = \begin{bmatrix} L_{11} & L_{12} \\ L_{21} & L_{22} \end{bmatrix} \qquad (3)$$

in $H_n$ such that the spectrum of $L$ is $\{E_1, \ldots, E_d, \ldots, E_n\}$, and the corresponding set of eigenvectors is $\Psi = \{\psi_\mu\}_1^n$. We shall let the indices i, j and $\mu$ run from 1 to $d$, from $d+1$ to $n$, and from 1 to $n$ respectively. Now the eigenequation

$$L\psi_\mu = E_\mu \psi_\mu \quad (\mu = 1, \ldots, n) \qquad (4)$$

is conveniently expressed in matrix form as

$$\begin{pmatrix} L_{11} & L_{12} \\ L_{21} & L_{22} \end{pmatrix} \begin{pmatrix} \alpha & a \\ 0 & b \end{pmatrix} = \begin{pmatrix} \alpha & a \\ 0 & b \end{pmatrix} \begin{pmatrix} E & 0 \\ 0 & e \end{pmatrix}, \qquad (5)$$

where

$\alpha = [\alpha_1 | \ldots | \alpha_d]$ is an $d \times d$ matrix whose columns are the vectors $\alpha_i$ $(i = 1, \ldots, d)$.

$\begin{bmatrix} a \\ b \end{bmatrix} = [\psi_{d+1} | \ldots | \psi_n]$ is an $n \times (n-d)$ matrix whose columns are the vectors

$\psi_j$ $(j = d+1, \ldots, n)$. This matrix is partitioned into an $d \times d$ matrix $a$ and an

$(n-d) \times d$ matrix $b$, so that $E\ \psi_j = \begin{pmatrix} a_j \\ b_j \end{pmatrix}, (j = d+1, \ldots, n)$. and $e$ are $(d \times d)$ and

$(n-d) \times (n-d)$ diagonal matrices with diagonal elements $E_i$ and $E_j$ respectively.

Now the eigenequation (5) is equivalent to tow sets of equations I an II
(i) $L_{11}\alpha = \alpha E$,     (ii) $L_{21}\alpha = 0$     (I)
(i) $L_{11}a + L_{12}b = ae$,     (ii) $L_{21}a + L_{22}b = be$     (II)

Since the set of vectors $\{\alpha_1, \ldots, \alpha_d\}$ is linearly independent, the second equation in (I), expanded as $L_{12}\alpha_i = 0 (i = 1, \ldots, d)$, shows that $L_{21}\beta = 0$ $(\forall \beta \in H_d)$, and hence $L_{21} = 0$. On substituting this result in the second set II, we obtain

(i) $L_{11}a + L_{12}b = ae$,     (ii) $L_{22}b = be$     (II)'

Now

$$\det \begin{pmatrix} \alpha & a \\ 0 & b \end{pmatrix} = \det \alpha . \det b \neq 0, \qquad (6)$$

because the set of columns of this matrix, namely $\{\psi_\mu\}_1^n$, is linearly independent. Since the set of vectors $\{\alpha_i\}_1^d$ is also linearly independent, $\det \alpha \neq 0$. Hence $\det b \neq 0$, and $b$ is invertible. In particular, the latter result affirms that no vector $b_j$ is zero, and that all vectors $b_j$ are eigenvectors of the operator $L_{22}$ belonging to the eigenvalues $E_j$. However, a vector $\begin{pmatrix} 0 \\ b_j \end{pmatrix}$ is not an eigenvector of $L$ unless $L_{12} = 0$, as it is implied by equation $(II)'$. Multiplying both equations in $(II)'$ from left by $b^{-1}$ we obtain

$$L_{22} = beb^{-1}, \quad L_{12}b = (ae - L_{11}a). \tag{7}$$

Hence

$$L = \begin{pmatrix} L_{11} & (ae - L_{11}a)b^{-1} \\ 0 & beb^{-1} \end{pmatrix}. \tag{8}$$

It is noted that the matrix which we seek, $L$, could have been directly obtained by the equation

$$L = \begin{pmatrix} \alpha & a \\ 0 & b \end{pmatrix} \begin{pmatrix} E & 0 \\ 0 & e \end{pmatrix} \begin{pmatrix} \alpha & a \\ 0 & b \end{pmatrix}^{-1}, \tag{9}$$

However this, in practice, is much more cumbersome than formula (8) which requires inversion of an $(n-d) \times (n-d)$ matrix, instead of inverting an $(n \times n)$ matrix as required by (9). In actual physical problems $n-d$ is much smaller than $d$. We shall see in the next section that, when the operator $L$ is Hermitian, a great deal of simplification could be achieved, and that the expression obtained does not involve matrix inversion at all.

## 3. Conditions of Diagonal Block Form

We discuss here the conditions imposed on the given eigenvectors so that the matrix $L$, which we have constructed, is immediately in a diagonal block form:

$$\widetilde{L} = \begin{bmatrix} L_{11} & 0 \\ 0 & beb^{-1} \end{bmatrix}. \tag{10}$$

For this to be the case we should have in (8) $L_{12} = 0$, which is equivalent to $L_{11}a = ae$, since $b$ is non-singular. This also could be seen from the first equation in (7), which shows that the operator $L_{12} : H_d^\perp \to H_d$ transforms each vector $b_j$ to a vector $(L_{11}b)_j = (ae - L_{11}a)_j = E_j a_j - L_{11}a_j$. An equation $L_{11}a_j - E_j a_j = 0$ holds if either of the following conditions hold:

(i) $a_j = 0$ or equivalently $\psi_j = \begin{pmatrix} 0 \\ b_j \end{pmatrix}$.

(ii) The eigenvector $\psi_j = \begin{pmatrix} a_j \\ b_j \end{pmatrix}$ with $a_j \neq 0$ belongs to an eigenvalue $E_j = E_i \in \{E_1,...,E_d\}$. In this case $a_j$ is an eigenvector of $L_{11}$ belonging also to the degenerate eigenvalue $E_j$, and hence $\psi_j - \psi_i = \begin{pmatrix} 0 \\ b_j \end{pmatrix}$ is also an eigenvector of $L$ belonging to the degenerate eigenvalue $E_j = E_i$.

We deduce accordingly that if for some $j \in [d+1, n]$, it is given that $a_j \neq 0$ and that $a_j$ is not an eigenvector of $L_{11}$ then $L$ cannot immediately assume a diagonal block form. In particular if the sets $\{E_i\}_1^d$ and $\{E_j\}_{d+1}^n$ do not intersect, then all $a_j$ have to vanish, if $L$ is to assume immediately a diagonal block form. On the other hand and if either of the above conditions (i) or (ii) hold, we may replace every vector $\psi_j$ given in the data by its projection on $\Pi^\perp$.

In all cases the block form (10) is equivalent, through a similarity transformation, to the diagonal block form (8). In fact it is easy to show that the similarity transformation effected by the matrix

$$T = \begin{pmatrix} 1 & -ab^{-1} \\ 0 & 1 \end{pmatrix}$$

has the following properties

1. It leaves all the vectors $\psi_i$ unaltered, i.e. $\tilde{\psi}_i \equiv T\psi_i = \psi_i$.
2. It projects every vector $\psi_j$ on the subspace $\Pi^\perp$, i.e. $\tilde{\psi}_j \equiv T\psi_j = \begin{pmatrix} 0 \\ b \end{pmatrix}$.
3. It transforms $L$ to $\tilde{L}$, i.e. $\tilde{L} = TLT^{-1}$.

## 4. Incorporation of New Information into the Hamiltonian

We mean by this title that given a Hermitian operator $H_{11}$ (for example the Hamiltonian of some physical system) as well as a set of numbers and a corresponding set of vectors, then our goal is to construct a new Hermitian operator $H$ whose spectrum is composed of the spectrum of $H_{11}$ as well as these given numbers, and whose corresponding eigenvectors consist of the set of eigenvectors of $H_{11}$ in addition to the given vectors. There are, of course, some conditions which are to be satisfied by the given vectors. These conditions will be spelled out in the sequel.

To state the problem concretely, let $H_{11}$ be a Hermitian operator in a Hilbert space $\Pi$, and assume that $H_{11}$ possesses the discrete spectrum $E_i$ $(i = 1,...,d)$ with the corresponding linearly independent eigenvectors $\psi_i$ $(i = 1,...,d)$, so that $H_{11}\psi_i = E_i\psi_i$ $(i = 1,...,d)$. Suppose also that we have a set of numbers $E_j$ $(j = d+1,...,n)$ and a set of orthogonal linearly independent vectors $\psi_j$ $(j = d+1,...,n)$. We seek to construct a Hermitian operator $H$ in $H_n$ such that

$$H\psi_\mu = E_\mu \psi_\mu \quad (\mu = 1,...,n) \tag{11}$$

Since the eigenvectors of a Hermitian operator that belong to different eigenvalues have to be orthogonal, and that which belong to the same eigenvalue can be orthogonalized by Gram-Schmidt procedure [4], we assume that the given vectors together with the eigenvectors of $H_{11}$ are orthogonal:

$$<\psi_\mu | \psi_\nu> = 0 \qquad (\mu \neq \nu; \mu, \nu = 1, ...., n) \qquad (12)$$

For notational convenience we shall set $\psi_\mu \equiv |\mu>$ and adopt Dirac's notations, so that the last equation takes the form $<\mu|\nu> = 0 (\mu \neq \nu)$. To gain insight into the method of solution, we consider as a first step the operator

$$L = H_{11} + <j|j>^{-1} \{E_j | j><j| - H_{11} | j><j|\} \qquad (13)$$

The operator $|j><j|$ with j fixed on one value in the range $\{d+1, ..., n\}$ is the projection operator on a one-dimensional subspace of $H_{d+1}$ generated by the vector $\psi_j \equiv |j>$. Evidently, all operators appearing in equation (13) act in the Hilbert space $H_{d+1}$. In particular $H_{11}$ is identified with the operator $\begin{bmatrix} H_{11} & 0 \\ 0 & 0 \end{bmatrix}$ in $H_{d+1}$. In notations more familiar to mathematicians, equation (13) is written thus

$$L = H_{11} + \|\psi_j\|^{-2} \{E_j P_j - H_{11} P_j\}$$

where $P_j$ is the projection operator on $\psi_j$. We shall show now that the spectrum of the operator $L$ is $\{E_1, ..., E_d, E_j\}$ and that the corresponding eigenvectors are $\{|1>, .... |d>, |j>\}$. Indeed

$$L|i> = H_{11}|i> + <j|j>^{-1} (E_j - H_{11})|j><j|i>$$
$$= E_i|i> + <j|j>^{-1} [(E_j - H_{11})|j>]<j|i>$$
$$= E_i|i>.$$
$$L|j> = H_{11}|j> + <j|j>^{-1} (E_j - H_{11})|j><j|j>$$
$$= H_{11}|j> + E_j|j> - H_{11}|j> = E_j|j>.$$

The generalization of the expression (13) to obtain an operator $H$ in $H_n$ that fulfill our requirements is straightforward. It can be checked easily that the following operator

$$H = H_{11}(I_n - \sum_{j=d+1}^{n} |j><j|) + \sum_{j=1+d}^{n} E_j |j><j| \qquad (14)$$

satisfies the eigenequation $H|\mu> = E_\mu |\mu>, (\mu = 1, ..., n)$. Indeed

$$(\mu \leq d) \Rightarrow H|\mu> = H_{11}(|\mu> - \sum_{i=d+1}^{n}|j><j|\mu>) + \sum_{j=d+1}^{n}E_j|j><j|\mu>$$

$$= H_{11}|\mu> = E_\mu|\mu>$$

$$(\mu > d) \Rightarrow H|\mu> = H_{11}(|\mu> - \sum_{j=d+1}^{n}|j><j|\mu>) + \sum_{j=d+1}^{n}E_j|j><j|\mu>$$

$$= H_{11}(|\mu> - \sum_{j=d+1}^{n}|j>\delta_{j\mu}) + \sum_{j=d+1}^{n}E_j|j>\delta_{j\mu}$$

$$= H_{11}(|\mu> - |\mu>) + E_\mu|\mu>$$

$$= E_\mu|\mu>.$$

It remains thus to prove that the operator $H$, which we have just constructed, is Hermitian.

## 5. On The Hermicity of $H$.

Denote by Q and P respectively the projection operators from $H_n$ onto $\Pi$ and $\Pi^\perp$, where $\Pi^\perp$ is the orthogonal complement of $\Pi$, then [4]

$$Q + P = I_n, \quad QP = 0, \quad Q^2 = Q, \quad P^2 = P \qquad (15)$$

In either basis $\{e_\mu\}_1^n$, or $\{\psi_\mu\}_1^n \equiv \{|\mu>\}_1^n$ in $H_n$ the projection operator Q is represented by the matrix

$$Q = \begin{bmatrix} I_d & 0 \\ 0 & 0_{n-d} \end{bmatrix}. \qquad (16)$$

Indeed

$$Q_{\mu\nu} = <\mu|Q|\nu> = 0 \text{ if } \mu > d \text{ or } \nu > d$$
$$= <i|Q|j> \quad \text{if } \mu = i \leq d \text{ and } \nu = j \leq d$$
$$= \delta_{ij}$$

In fact the representation (16) is always valid as long as every basis element does not have components in both $\Pi$ and $\Pi^\perp$. This is evident since the restriction of $Q$ to $\Pi$ is merely the identity operator. Similarly, the matrix of the projection operator $P$ is

$$P = \begin{bmatrix} 0_d & 0 \\ 0 & I_{n-d} \end{bmatrix} \qquad (17)$$

The equation

$$P = \sum_{j=d+1}^{n}|j><j| \qquad (18)$$

is valid since both of its sides are projection operators on $\Pi^\perp$. The operator $H$ can therefore be expressed thus:

$$H = H_{11}(I_n - P) + \sum_{j=d+1}^{n}E_jP_j = H_{11}Q + \sum_{j=d+1}^{n}E_jP_j = QH_{11}Q + \sum_{j=d+1}^{n}E_jP_j.$$

Recalling that projection operators are Hermitian, we deduce that $H = H^+$, i.e. the operator $H$ is equal to its adjoint operator $H^+$, and hence is Hermitian.